\documentclass[aps,prl,floats,twocolumn,superscriptaddress,showpacs]{revtex4}

\usepackage{epsfig}
\usepackage{latexsym,amsmath}

\newcommand{\avk}[1]{\langle {#1}\rangle}

\newcommand{\nper}{{\nu_\perp}}

\newcommand{\npar}{{\nu_\parallel}}

\newcommand{\rhoa}{\bar{\rho}_a}

\begin{document}

\title{Non mean-field behavior of the contact process on scale-free networks}

\author{Claudio Castellano}
%\affiliation{INFM Unit\`a Roma 1 and SMC and
%Dipartimento di Fisica, Universit\`a di Roma ``La Sapienza'',
%5P.le Aldo Moro 2, I-00185 Roma, Italy \\
%and Istituto dei Sistemi Complessi, CNR,
%Via Dei Taurini 9, 00185 Roma, Italy}
\affiliation{CNR-INFM, SMC, Dipartimento di Fisica, Universit\`a di Roma
``La Sapienza'', P.le Aldo Moro 2, I-00185 Roma, Italy and
CNR, Istituto dei Sistemi Complessi, Roma, Italy}

\author{Romualdo Pastor-Satorras}
\affiliation{Departament de F\'\i sica i Enginyeria Nuclear, Universitat
  Polit\`ecnica de Catalunya, Campus Nord B4, 08034 Barcelona, Spain}

\date{\today}

\begin{abstract}
  We present an analysis of the classical contact process on
  scale-free networks. A mean-field study, both for finite and
  infinite network sizes, yields an absorbing-state phase transition
  at a finite critical value of the control parameter, characterized
  by a set of exponents depending on the network structure.  Since
  finite size effects are large and the infinite network limit cannot
  be reached in practice, a numerical study of the transition requires
  the application of finite size scaling theory.  Contrary to other
  critical phenomena studied previously, the contact process in
  scale-free networks exhibits a non-trivial critical behavior that
  cannot be quantitatively accounted for by mean-field theory.
\end{abstract}

\pacs{89.75.-k,  87.23.Ge, 05.70.Ln}

\maketitle

Recent years have witnessed an upsurge in the study of the structure
and function of complex networks 
%occurring in natural, sociological,
%and technological contexts 
\cite{barabasi02}.  The interest
in these systems has been driven by two motivations.  First,
a large amount of work has been devoted to the structural
characterization of real networks, the main contribution being the
discovery of their seemingly ubiquitous scale-free (SF) nature
\cite{barab99}, characterized by a probability $P(k)$ that an element
in the network (a \textit{vertex}) is connected to other $k$ elements
(has \textit{degree} $k$), scaling as a power-law, $P(k) \sim k^{-\gamma}$,
with a certain degree exponent $\gamma$ usually in the range $2 < \gamma <3$
\cite{barab99}.  On the other hand, a great deal of attention has been
drawn by the profound effects that a heterogeneous SF
connectivity pattern can have on dynamical processes
on complex networks. In particular, models exhibiting critical phase
transitions such as ferromagnets \cite{isingvespi},
percolation \cite{havlin00} or epidemic models
\cite{pv01a} have been studied on this class of networks.
Present understanding of critical behavior in complex networks is
essentially based on the application of mean-field (MF) techniques
\cite{stanley}.  In the study of critical phenomena on regular
substrates (lattices), MF theory plays a prominent role.  Despite its
simplicity, it describes qualitatively most phase transitions.
%, both in and out of equilibrium.  
Moreover, while in low spatial dimensions
fluctuations cannot be neglected, above the upper critical dimension
MF is even quantitatively correct, providing exact expressions for
critical exponents. In the case of complex networks, since they
usually fulfill the so-called \textit{small-world} (SW) property
\cite{watts98}, 
such systems can be viewed as infinite
dimensional, and therefore one can expect MF to provide
an exact description of dynamical processes taking place on
networks, once degree heterogeneity has been taken into account.
Results obtained so far seem to validate this idea,
particularly for SF networks \cite{gdm03}.

In this paper we show the inadequacy of MF theory to describe some
dynamical processes on SF networks, by considering 
%one of the simplest
%non-equilibrium critical phenomena: 
the absorbing-state phase
transition of the contact process (CP) \cite{harris74}.  We develop a
MF theory for the CP on generic networks, and solve it for SF
networks, obtaining expressions for both infinite and finite networks.
These expressions are compared with numerical simulations of the CP on
%different classes of 
SF networks, which are analyzed applying a finite
size scaling (FSS) theory \cite{marro99}. Our results show
that the the critical exponents of CP on SF networks
cannot be accounted for by MF theory, and signal the presence
of nontrivial fluctuation effects.

We consider the classical CP on generic networks, defined as follows
\cite{marro99}: An initial fraction $\rho_0$ of vertices is randomly
chosen and occupied by a particle.  The time evolution of the process
runs as follows: At each time step $t$, a particle in a vertex $v_i$
is chosen at random. With probability $p$, this particle disappears.
With probability $1-p$, on the other hand, the particle may generate
an offspring. To do so, a vertex $v_j$, nearest neighbor of the vertex
$v_i$, is randomly chosen.  If vertex $v_j$ is occupied by a particle,
nothing happens; if it is empty, a new particle is created on $v_j$.
In any case, time is updated as $t \to t+1/n(t)$, where $n(t)$ is the
number of particles present at the beginning of the time step
%, so that
%each particle is selected on average once during a time unit
\cite{marro99}.  In Euclidean $d$-dimensional lattices, the CP
undergoes a continuous transition
%, located 
at a critical point $p_c$,
separating an absorbing phase from an active one~\cite{marro99}.  This
phase transition is characterized in terms of the order parameter $\rho$,
defined as the average density of particles in the steady state.  For
$p>p_c$, an absorbing phase with $\rho=0$ is observed. For $p<p_c$
%, on the other hand, 
the system sets in an active phase with a nonzero
order parameter, obeying $\rho \sim \Delta^\beta$ for $p \to p_c^{-}$,
where $\Delta = p_c-p$. Close to $p_c$, the system is further characterized by
diverging correlation length and time scales, namely $\xi \sim
|\Delta|^{-\nper}$ and $\tau \sim |\Delta|^{-\npar}$.
The critical exponents $\beta$,
$\nper$ and $\npar$ provide a full characterization of the phase
transition of CP in Euclidean lattices.  Below the critical dimension
$d_c=4$, the exponents are nontrivial, and depend explicitly on $d$.
For $d>d_c$, the exponents take the classical MF values $\beta=\npar=1$,
$\nper=1/2$.

In order to gain analytical insight on the behavior of the CP
on complex networks, we consider as usual a MF rate equation for the
partial densities $\rho_k(t)$ of occupied vertices of degree $k$
\cite{pv01a}, from which the total density is obtained as $\rho(t)
= \sum_k \rho_k (t) P(k)$. The rate equation for the partial densities in a
network characterized by a degree distribution $P(k)$ and degree
correlations given by the conditional probability $P(k'|k)$ that a
vertex of degree $k$ is connected to a vertex of degree $k'$
\cite{marianprocpaper} can be thus written as \cite{michelediffusion}
\begin{equation}
  \partial_t  \rho_k(t) = -  \rho_k(t) +  \lambda  k [1-\rho_k(t)] \sum_{k'} \frac{P(k' |
    k) \rho_{k'} (t)}{k'}.
  \label{eq:1}
\end{equation}
The destruction term, due to the vanishing of particles---with
probability $p$---at vertices of degree $k$, is proportional to the
density of occupied vertices of degree $k$, $\rho_k(t)$. The creation
term, on the other hand, is due to the generation of offsprings from
particles located at nearest neighbor vertices of empty vertices of
degree $k$. Thus, it is proportional to the probability
$[1-\rho_k(t)]$ that a vertex of degree $k$ is empty, that it is
connected to a vertex of degree $k'$, $P(k' | k)$, that this one is
occupied, $\rho_{k'}(t)$, that it generates an offspring, $1-p$, and that
this neighbor chooses to create the offspring on the vertex of degree
$k$ under consideration, $(1/k')$.  In Eq.~(\ref{eq:1}) we have
performed a rescaling of time, introducing the new parameter $\lambda=
(1-p)/p$.  For uncorrelated networks, where $P(k' | k) = k' P(k') /
\avk{k}$ \cite{marianprocpaper}, the rate equations take the form
\begin{equation}
  \partial_t  \rho_k(t) = - \rho_k(t) +
  \lambda \frac{k}{\avk{k}} [1-\rho_k(t)] \rho(t).
  \label{eq:2}
\end{equation}

The solution of Eq.~(\ref{eq:2}) depends on the nature of the degree
distribution. For \textit{homogeneous} networks, in which $P(k)$
decays exponentially for large degrees, all vertices are approximately
equivalent.
Therefore, we have $k \simeq \langle k \rangle$ and
$\rho_k(t) \simeq \rho(t)$, so that the rate equation reads
$ \partial_t \rho(t) = - \lambda \rho^2(t) + \tilde{\Delta}
\rho(t)$, where $\tilde{\Delta}=\lambda -1$.
This equation describes the behavior
of the CP in an infinite dimensional system; the critical point
$\lambda_c=1$ ($p_c=1/2$) and the exponent $\beta=1$ are thus immediately found
\cite{marro99}.  Moreover, from the solution
$\rho(t) \sim \tilde{\Delta}/ \lambda -
e^{- \tilde{\Delta} t}$, we can identify the relevant characteristic time
$\tau \sim \tilde{\Delta}^{-1}$ and the corresponding exponent $\npar =1$.

For \textit{heterogeneous} networks, with a degree distribution
exhibiting large fluctuations, the solution of Eq.~(\ref{eq:2})
must consider explicitly the $k$ dependence of the partial densities.
Information on the active phase is obtained by imposing the
steady state condition, $\partial_t \rho_k(t)=0$, which
yields the nonzero solutions
\begin{equation}
  \rho_k = \frac{\lambda k \rho / \avk{k}}{1+\lambda k \rho / \avk{k}},
  \label{eq:4}
\end{equation}
where $\rho_k$ is now independent of time.  In order to determine the
critical behavior of the average density $\rho$ as a function of the
control parameter $\lambda$ we focus on the self-consistent equation for the
order parameter $\rho$.  By combining Eq.~(\ref{eq:4}) with the
expression for $\rho$, one obtains
\begin{equation}
\rho =  \frac{\lambda \rho }{\avk{k}} \sum_k  \frac{k P(k)}{1+\lambda k \rho / \avk{k}}. 
  \label{eq:6}
\end{equation}
This equation depends on the full degree distribution.  In the case of
SF networks, for which the degree distribution in the continuous
degree approximation is given by
$P(k) = (\gamma-1) m^{\gamma-1} k^{-\gamma}$, with
$m$ the minimum degree in the network, the solution will depend on the
degree exponent $\gamma$. Substituting the summation by an integral in
Eq.~(\ref{eq:6}), we obtain in the \textit{infinite network size
  limit} (i.e.  when the degree belongs to the range $[m, \infty]$) the
expression $\rho=F[1,\gamma-1,\gamma,-(\lambda \rho m/\avk{k})^{-1}]$,
where $F[a,b,c, z]$
is the Gauss hyper-geometric function~\cite{abramovitz}.  To evaluate
the critical behavior for small $\rho$, we invert this expression using
the asymptotic expansion of the hyper-geometric function for $z \to -\infty$,
obtaining~\footnote{The case $\gamma=3$ gives logarithmic
corrections that will be discussed elsewhere.}
$\rho(\lambda) \sim (\lambda-1)^{1/(\gamma-2)}$ for $2 <\gamma < 3$
and $\rho(\lambda) \sim \lambda-1$ for $\gamma > 3$.
Hence the MF solution in infinite networks gives the critical exponent
$\beta^\mathrm{MF}=1/(\gamma-2)$ for $\gamma<3$, and $\beta^\mathrm{MF}=1$
for $\gamma> 3$, while the critical threshold is $\lambda^\mathrm{MF}_c=1$
($p_c^\mathrm{MF}=1/2$), independent of the degree exponent.

The previous results, corresponding to networks of infinite size, are
however strongly affected by finite size effects. 
The difference between an infinite SF network and one of finite
size $N$ is that the latter has a cut-off or maximum degree $k_c$,
which is a function of the network size \cite{dorogorev}, and that,
for uncorrelated
networks, scales as $k_c(N) \sim N^{1/2}$ \cite{mariancutofss}.  The
presence of the cut-off restricts the possible values of the degree
to the range $[m, k_c]$.  For
\textit{finite networks}, Eq.~(\ref{eq:6}) yields, in the continuous
degree approximation,
$\rho=F[1,\gamma-1,\gamma,-(\lambda \rho m/\avk{k})^{-1}] - (k_c
/m)^{1-\gamma} F[1,\gamma-1,\gamma,-(\lambda \rho k_c/\avk{k})^{-1}]$.
To obtain the MF critical behavior for a finite network, we evaluate
this expression in the limit $\rho \to 0$ for generic $\rho k_c$, obtaining
\begin{equation}
  \rho = \frac{\avk{k}}{\lambda k_c}
  f\left(\frac{\avk{k} \tilde{\Delta} k_c^{\gamma-2}}{\lambda m^{\gamma-1}}
  \right),
\label{eq:3}
\end{equation}
where $f(x)$ is the inverse of the function $g(x) = (F[1,\gamma-1,\gamma,-1/x]+
\Gamma(\gamma) \Gamma(2-\gamma) x^{\gamma-1})/x$.  For $x \gg 1$, $f(x)$
diverges as
$x^{1/(\gamma-2)}$, in agreement with the infinite size limit $\rho \sim
\tilde{\Delta}^{\beta^\mathrm{MF}}$.  For small arguments, the function
$f(x)$ vanishes for $x \to x_0^+$ with $x_0>0$. Therefore, for finite
$N$, if $\Delta$ is made smaller than a value proportional to $k_c^{2-\gamma}$,
then $\rho=0$. This indicates that the effective critical point is
shifted in the active phase in finite networks and that the
convergence to the infinite network limit is slow, since corrections
vanish as $N^{(2-\gamma)/2}$, i.e., with an exponent smaller than $1$.

>From Eq.~(\ref{eq:3}) we can ascertain when, in a system of $N$
vertices, finite size effects start to play a relevant role.  It is
clear that for $x <1$, i.e.
$\tilde{\Delta} < \tilde{\Delta}_\times \sim (k_c/m)^{2-\gamma}$
such effects start to appear, so that one expects deviations from the
decay $\rho \sim \tilde{\Delta}^{\beta^\mathrm{MF}}$.  In terms of the order
parameter this implies that finite size effects are negligible only if
$\rho \gg \rho_\times = \avk{k} /(\lambda k_c)$. 
This condition introduces serious
constraints on the precision of the numerical estimate of the critical
point 
%position, 
when performed by measuring the dependence of the $\rho$
on $p$ in the active phase.

This difficulty %in the numerical study of the CP in SF networks
can be overcome by using the finite size scaling (FSS) technique
\cite{marro99,zanette01:_critic}. The FSS
technique is based on the observation that, even below the critical
point, the density of active sites in surviving runs $\rhoa$
reaches a quasi-steady state whose average is a decreasing function of
the network size $N$ \cite{marro99}. The FSS ansatz assumes
the dependence of $\rhoa$ as a function of $\Delta=p_c-p$ and $N$ 
\begin{equation}
  \rhoa(\Delta,N) = N^{-\beta/ \nu_\perp} \;
  f(\Delta N^{1/ \nu_\perp}),
  \label{eq:9}
\end{equation}
where the scaling function has the asymptotic behavior $f(x) \to x^{\beta}$
for $x\to\infty$, and $f(x) \to \mathrm{const.}$ for $x\to0$. 
Compatibility with the theoretical behavior given in Eq.~(\ref{eq:3})
for $x\gg 1$ requires that the exponents assume the MF values
$\beta=1/(\gamma-2)$ and $\nu_\perp=2/(\gamma-2)$
in uncorrelated networks with cut-off scaling as $N^{1/2}$.  Thus, at
$p_c$ the function $\rhoa$ is expected to decay as
\begin{equation}
  \rhoa(0, N) \sim N^{-\beta/ \nu_\perp}.
  \label{eq:5}
\end{equation}
This fact can be used to estimate the critical point, as the value of
$p$ yielding a power-law behavior for $\rhoa$ as a function of
$N$. MF theory predicts in this case an exponent $\beta/ \nu_\perp = 1/2$.

In order to check these results, we have simulated numerically the
CP on the uncorrelated configuration model (UCM)~\cite{ucmmodel},
which generates uncorrelated SF networks with
arbitrary degree exponent and cut-off $k_c(N) \sim N^{1/2}$.  We consider
the particular cases $\gamma=2.75, 2.5$ and $2.25$.  Simulations were
performed with at least 100 runs of the CP for each realization of the network
and no less than 100 such realizations for each value of $\gamma$ and $N$.

\begin{figure}[t]
  \epsfig{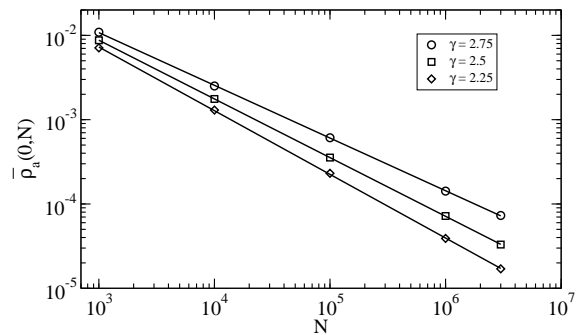}
  \caption{Scaling of $\rhoa(0 , N)$ as a function of the network size 
    for the CP on UCM networks for three values of $\gamma$. The critical
    points and exponents $\beta/ \nu_\perp$ obtained are reported in
    Table~\ref{tab:exponents}.  Solid lines are least squares fits.}
  \label{fig:fss_crit}
\end{figure}

We focus first on the determination of the critical point and of the
exponent ratio $\beta/ \nu_\perp$ via Eq.~(\ref{eq:5}).
\begin{figure}[t]
  \epsfig{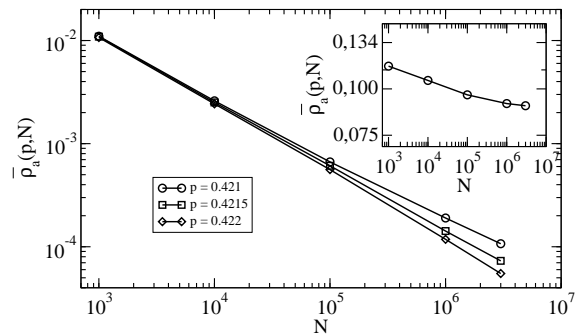}
  \caption{$\rhoa(p , N)$ as a function of the network size $N$
    for the CP on UCM networks with $\gamma=2.75$, for values of $p$
    close to $p_c$. Solid lines are guides to the eye.
    Inset: $\rhoa(p , N)$ for $p=0.36$, showing that the infinite
    $N$ limit is not reached even far from $p_c$.}
  \label{fig:fss}
\end{figure}
In Fig.~\ref{fig:fss} we display the density of active sites in
surviving runs $\rhoa$ as a function of the network size, for $\gamma=2.75$
and several values of the control parameter $p$.  The critical point
is identified as the value of $p$ such that the decay of $\rhoa$ vs
$N$ is a pure power-law, which for this particular value of $\gamma$
corresponds to $p_c=0.4215$.  In Fig.~\ref{fig:fss_crit} we show the
decay of $\rho_a$ vs $N$ at criticality, for the three values of $\gamma$
considered.  In all cases a very clear power-law behavior is
identified, yielding the exponent ratios $\beta/\nu_\perp=0.63$ for
$\gamma=2.75$, 0.70 for $\gamma=2.50$, and 0.76 for $\gamma=2.25$.
Notice that these values are in strong disagreement with the MF
prediction $\beta/\nu_\perp=1/2$.

In principle the exponent $\beta$ should be determined by the way the
steady state density of particles vanishes as a function of $p_c-p$ in
the limit $N \to \infty$.  However, it turns out that, even far from $p_c$,
it is numerically impossible to fulfill the condition $\rho \gg \rho_\times$ that
guarantees the absence of finite size effects.  This is apparent in
the inset of Fig.~\ref{fig:fss}, where one can see that even for
$p=0.36$ the value of $\rhoa$ slightly depends on $N$ even for large
sizes.  This very slow convergence 
%to the asymptotic $N$-independent values 
hampers the determination of $\beta$ in the usual way.

\begin{figure}[t]
  \epsfig{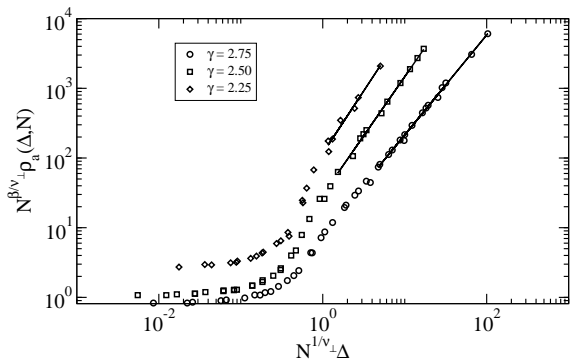}
  \caption{Data collapse of  $N^{\beta/\nu_\perp} \rhoa(\Delta , N)$  as a
    function of $\Delta N^{1/ \nu_\perp}$ for the CP on  UCM networks.
    Data are for system size ranging from $N=10^3$ to $N=10^7$.
    For clarity the data for $\gamma=2.25$ have been shifted upwards.
    %The values of $\beta/\nu_\perp$ are obtained from Fig.~\ref{fig:fss_crit}, 
    The values of $\nu_\perp$ are those giving the best data collapses.
    The straight lines are least squares fits to the
    behavior $f(x) \to x^{\beta'}$, for large $x$.}
  \label{fig:collapse}
\end{figure}

The numerical value of $\beta$ and $\nu_\perp$ can be however
determined via the FSS analysis, by noticing that from Eq.~(\ref{eq:9})
we expect data for the surviving density to collapse onto a universal
function $f(x)$ when plotting $N^{\beta/ \nu_\perp} \rhoa(\Delta , N)$
as a function of $\Delta N^{1/\nu_\perp}$.
Using the values of $\beta/\nu_\perp$ obtained from
Fig.~\ref{fig:fss} we try the collapse for various values of
$\beta$ (and, as a consequence, of $\nu_\perp$).
The best data collapses, shown in Fig.~\ref{fig:collapse},
are obtained for the values of $\beta$ and $\nu_\perp$ reported
in Table~\ref{tab:exponents}.
The scaling function $f(x)$ is expected to diverge
as $x^\beta$ for asymptotically large $x$.
We measure how the numerical scaling functions go for large $x$
in Fig.~\ref{fig:fss_crit}, finding the effective behavior
$f(x) \sim x^{\beta'}$ with $\beta'=1.43$, $1.70$, and $1.79$.
The disagreement with the values of $\beta$ providing the
best data collapse indicates that the values of $x$ that can be reached
are not large enough to see the truly asymptotic $x^\beta$ behavior.

\begin{table}[b]
  \caption{
    Critical points and critical exponents for the CP on  UCM networks,
    compared with MF values.
    Numbers in parentheses indicate the uncertainty in
    the last digit.}
  \label{tab:exponents}
  \begin{ruledtabular} 
    \begin{tabular}{l|cccc}
              & $p_c$   & $\beta/\nu_\perp $ & $\beta$   & $\nu_\perp$ \\\hline
MF            & $1/2$       & $1/2$     & $1/(\gamma-2)$ & $2/(\gamma-2)$   \\
$\gamma=2.75$ & $0.4215(5)$ & $0.63(4)$ & $1.52(5)$      & $2.4(2)$ \\
$\gamma=2.50$ & $0.4425(5)$ & $0.70(3)$ & $2.0 (1)$      & $2.8(3)$ \\
$\gamma=2.25$ & $0.465(3) $ & $0.76(5)$ & $2.4 (1)$      & $3.2(3)$

    \end{tabular} 
  \end{ruledtabular}
\end{table}

\begin{figure}
  \epsfig{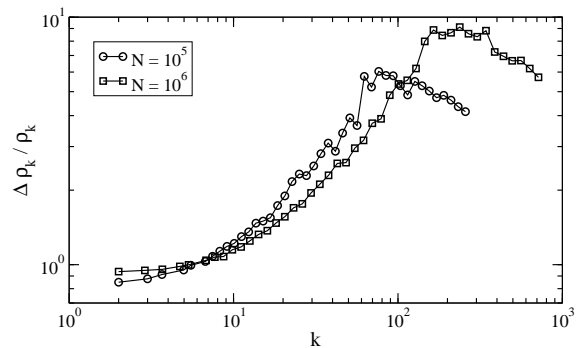}
  \caption{Ratio of the standard deviation $\Delta \rho_k$ 
of the distribution of the density of occupied vertices of degree $k$
over the average $\rho_k$, versus the degree $k$, for $\gamma=2.75$,
$p=p_c=0.4215$ and two values of the system size $N$.}
  \label{fluct}
\end{figure}

We summarize the results obtained in
Table~\ref{tab:exponents}.
%, where they are compared with the
%corresponding MF values. 
The conclusion of this analysis is that the
CP in SF networks exhibits an absorbing-state phase
transition at a non-zero critical point, whose exponents
are not correctly predicted by MF
theory~\footnote{We have checked that for complete and random
graphs MF predictions are instead verified}.
This result is surprising: it is generally believed
that dynamical models on SF networks find in this formalism a
correct description, once degree heterogeneity has been taken into
account~\cite{pv01a}.
The results presented here open two clear perspectives for future work.
The first is the quest for a theory beyond MF for the CP on heterogeneous
networks. 
The second is the investigation of the limits of MF theory for generic
dynamical processes on heterogeneous topologies.
%, since many other
%models find in this formalism a correct description.
A useful indication along these lines is provided by
Fig.~\ref{fluct}, where we plot the ratio of the standard deviation
$\Delta \rho_k$ over the mean density $\rho_k$ restricted to vertices
of degree $k$, at the critical point.
It turns out that the fluctuations are much larger than the average
except for the smallest $k$, all the more so for increasing system
size.  Hence it is incorrect to assume (as MF theory does) that the
mean $\rho_k$ provide all information about the distribution of densities
for vertices of a certain degree.  The inclusion of these large
fluctuations and the understanding of their role are thus key points
to develop a theory beyond MF, able to fully capture the physics of
collective phenomena on SF networks.

\begin{acknowledgments}
  We thank Alessandro Vespignani for useful discussions.
  R.P.-S. acknowledges financial support from the Spanish MEC (FEDER),
  under project No.  FIS2004-05923-C02-01 and additional support from
  the MCyT and DURSI (Spain). C.C. acknowledges financial support from
  Azione Integrata Italia-Spagna IT1792/2004.
\end{acknowledgments}

\end{document}